# Evaluating Research Quality with Large Language Models: An Analysis of ChatGPT's Effectiveness with Different Settings and Inputs

Mike Thelwall, Information School, University of Sheffield, UK.

**Purpose**: Evaluating the quality of academic journal articles is a time consuming but critical task for national research evaluation exercises, appointments and promotion. It is therefore important to investigate whether Large Language Models (LLMs) can play a role in this process.

**Design/methodology/approach**: This article assesses which ChatGPT inputs (full text without tables, figures and references; title and abstract; title only) produce better quality score estimates, and the extent to which scores are affected by ChatGPT models and system prompts.

**Findings**: The optimal input is the article title and abstract, with average ChatGPT scores based on these (30 iterations on a dataset of 51 papers) correlating at 0.67 with human scores, the highest ever reported. ChatGPT 4o is slightly better than 3.5-turbo (0.66), and 4o-mini (0.66).

**Research limitations**: The data is a convenience sample of the work of a single author, it only includes one field, and the scores are self-evaluations.

**Practical implications**: The results suggest that article full texts might confuse LLM research quality evaluations, even though complex system instructions for the task are more effective than simple ones. Thus, whilst abstracts contain insufficient information for a thorough assessment of rigour, they may contain strong pointers about originality and significance. Finally, linear regression can be used to convert the model scores into the human scale scores, which is 31% more accurate than guessing.

**Originality/value**: This is the first systematic comparison of the impact of different prompts, parameters and inputs for ChatGPT research quality evaluations.

## 1 Introduction

Evaluating the quality of academic research is necessary for many national research evaluation exercises, such as in UK and New Zealand (Buckle & Creedy, 2024; Sivertsen, 2017), as well as for appointments, promotions, and tenure (Pontika et al., 2022; Tierney & Bensimon, 1996). These evaluations are expensive and reduce the time that academics can spend conducting research, so automated alternatives or support are desirable. The need for shortcuts has given rise to the field of bibliometrics/scientometrics, which has (controversially: MacRoberts & MacRoberts, 2018) developed a wide range of academic impact or quality indicators, typically based on citation counts (Moed, 2006). Attempts to use bibliometric information to directly score journal articles for research quality have obtained mixed results, however, with article level Pearson correlations between traditional machine learning (e.g., extreme gradient boost) predictions varying between 0.028 (Art and Design) and 0.562 (Clinical Medicine) for REF2021 (Thelwall et al., 2023). It is therefore logical to assess whether LLMs can be more accurate at predicting quality scores, given that they are relatively accurate at a wide range of language processing tasks (Bang et al., 2023; Kocoń et al., 2023), and have been proposed for scientometric purposes (Bornmann & Lepori, 2024). In fact, one funder is now using LLMs to support the initial triage of biomedical grant proposals, creating a pool of apparently weaker submissions for human reviewers to consider for rejection before the full peer review process. The experimental results showed that LLMs were not reliable enough to be used without human judgements but had an above random chance of identifying weak submissions and could save time by flagging these for human triage (Carbonell Cortés, 2024).

Many previous studies have assessed the extent to which LLMs can give useful *feedback* on academic works, usually to help reviewers/editors analysing conference/journal submissions or to support authors with text revisions. These have tended to show that LLMs can make useful suggestions, and their comments may even extensively overlap with those of human reviewers (Liang et al., 2024b). The main current challenge is that LLM-generated reviews are plausible enough that busy reviewers may cheat by using them. Because of this threat to the integrity of the peer review process and for copyright infringement, some publishers, like Elsevier (2024), instruct reviewers and editors not to use LLMs. Alternative approaches include detecting LLM-authored reviews (Liang et al., 2024a) or harnessing LLM outputs to build a reward system for reviewers creating high quality reports (Lu et al., 2024).

Whilst the extent to which LLMs can make useful comments on manuscripts is now supported by substantial evidence, their ability to make overall quality judgements for unreviewed and reviewed papers is less clear. The two tasks are different, albeit overlapping, because a pre-publication review might focus primarily on checking rigour and finding errors, also considering whether an article meets thresholds for originality and significance. In contrast, post-publication quality review might focus on the broader context, such as the extent to which the rigour, originality, significance combine to make a useful scholarly contribution.

For post-publication research quality assessment, ChatGPT 4 can identify two characteristics of papers (novelty and engagement; accessibility and understandability) that correlate positively with citation counts (de Winter, 2024). Similar to the current paper, one study analysed unreviewed conference papers from the PeerRead dataset of 427 reviews of International Conference on Learning Representations 2017 submissions (Kang et al., 2018). It used the prompt, "You are a professional reviewer in computer science and machine learning. Based on the given review, you need to predict the review score in several aspects. Choose a score from [1,2,3,4,5], higher score means better paper quality" with ChatGPT 2.5 or ChatGPT 3.5-16k, probably with the ChatGPT Applications Programming Interface (API). It found that reviewer scores of 1-5 weakly correlated with ChatGPT 3.5 predictions when only submission abstracts were entered (Spearman: 0.282) but not when full texts were entered (Spearman: 0.091) (Zhou et al., 2024). It is not clear if multiple reviewer scores were aggregated for the same paper, however. For (mainly) reviewed papers, one previous study of 51 information science papers from a single author (me) asked ChatGPT 4 (web interface) to score articles (full text PDFs or Word documents) for research quality using the instructions given to assessors in the UK Research Excellence Framework (REF) 2021. It elicited a score of 1*,2*,3* or 4* for each paper, repeating the process 15 times. Whilst the ChatGPT 4 scores correlated weakly (0.20) with the author's self-evaluations, when the 15 scores for each article were averaged, the correlation rose to 0.51 (Thelwall, 2024). The article did not compare different LLMs, inputs, or prompts.

The current article addresses a few gaps left by the above research. It seeks to verify and extend the prior suggestion that analysing abstracts gives better results than analysing full texts (Zhou et al., 2024). The focus is on ChatGPT rather than alternatives because ChatGPT has been shown to perform well on a wide variety of text processing tasks, as reviewed above, and there is insufficient evidence yet to prefer Google Gemini (e.g., Buscemi & Proverbio, 2024). This article also specifically investigates the relationship between the number of ChatGPT iterations and the accuracy of the prediction, extending a prior study that only compared 1 and 15 iterations (Thelwall, 2024). It also compares different ChatGPT models for research quality evaluation for the first time. Also for the first time, it directly assesses the accuracy of the scores and compares different system prompts. The overall purpose is to gain more insights into how to obtain the most accurate estimates from ChatGPT.

- RQ1: What is the optimal input for ChatGPT post-publication research quality assessment: full text, abstract, or title only?
- RQ2: What is the relationship between the number of ChatGPT iterations averaged, and the usefulness (correlation with human judgement) of its predictions?
- RQ3: Does ChatGPT model choice affect post-publication research quality assessment?
- RQ4: Are simpler system prompts more effective than complex system prompts?

## 2 Methods

The overall research design was to run a series of experiments guided by the research questions on a set of 51 articles with quality scores, using the ChatGPT API environment. For each experiment, the ChatGPT completion requests were carried out consecutively and then repeated a further 29 times to give 30 scores for each article. This seems like a large enough number (and double the previously published maximum) to reveal the relationship between the number of iterations and the usefulness of the average score predictions. The API environment is necessary to run the high volume of tests for the proposed experiments. It also has the added advantage that ChatGPT promises not to use data submitted within the API for training its models ("We do not train our models on inputs and outputs through our API", OpenAI, 2024), so each score should be independent of all the others. The queries were all submitted in July 2024.

### 2.1 Data

The raw data for this paper is a set of 51 information science journal articles that have either been published or prepared for submission and subsequently rejected or not submitted. All were written by the author, who has copyright, and were scored by him using the REF quality scale of 1*, 2*, 3* or 4* (REF, 2019), with which he is familiar. Papers were given a mid-score (e.g., 3.5*) if they seemed to fall between two score levels. The scores given to these papers have never been disclosed to anyone else or uploaded to any Artificial Intelligence (AI) system. This dataset is not ideal since (a) it is part of a single author's output and therefore not representative even of its field, (b) the author's scores are less relevant than the scores of more independent and less expert (on this topic) senior researchers, who would be the ones forming the evaluations in the most important context (e.g., the REF or promotion/appointment). Nevertheless, the papers are relatively homogeneous in topic, giving more scope for AI to differentiate quality differences from topic differences.

The REF quality criteria encompass the core dimensions of rigour, significance and originality (Langfeldt et al., 2020). Whilst the specific instructions are unique, the goals are nevertheless general in this sense. Quality criteria can differ between contexts, however, for example by emphasising local research needs (e.g., Barrere, 2020).

All articles were available in PDF format or as Word Documents. The Word documents were converted to text with Word's Save As text feature. The PDF documents were converted to text with PyMuPDF in Python (Convert_academic_pdf_to_jsonl.py in https://github.com/MikeThelwall/Python_misc). PyMuPDF discards images and converts the text into blocks, each of which is typically a paragraph or sentence. It guesses paragraphs, frequently making mistakes, and does not attempt to format tables. A program was written to clean the results, merging sentences into paragraphs, when appropriate, and removing page headers and footers (in the Webometric Analyst program free online at https://github.com/MikeThelwall/Webometric_Analyst: Services|PDF text file processing|Clean Fitz/PyMuPDF converted PDF text files from headers). These texts were then manually checked for paragraph structure and additional headers/footers, requiring extensive cleaning in each case. They were then processed to convert into Jsonl format for input into the ChatGPT API (Webometric Analyst: Services|Convert set of texts (e.g., cleaned PDFs) in folder to single Jsonl file). This generates less rich representations than the original PDFs/Word documents previously used (Thelwall, 2024) due to the lack of images and appropriately formatted tables and formulae, but has the advantage of being better structured in terms of paragraphs.

Several different datasets were generated to try different extents of input.

**Truncated dataset**: This consisted of the full text files without the reference list (not strongly relevant), the contents of tables (difficult to process by an LLM), the authors, and the keywords. This was designed to contain all the key information needed for an evaluation, given that core information in tables and figures would be referred to in the full text and the reference list is not essential to an evaluation, even though it can sometimes be statistically helpful in citation prediction (Kousha & Thelwall, 2024; Qiu & Han, 2024). Thus, in the spirit of document simplification (Zhou et al., 2024) it is probably a better input than a full text PDF, which the API would not accept anyway.

**Abstract dataset**: This consisted of the title and abstract alone after removing the authors, keywords and the remaining text.

**Title dataset**: This consisted of article titles alone.

## 2.2 GPT prompt setup

In machine learning it is typical to use separate development and training sets to allow an AI system to be configured with data that it is not tested on. This was not done in this case although there was a configuration exercise. This exercise consisted: of (a) fruitless tests with different prompts to try to get the score prediction to be reported more consistently, and (b) fruitless experiments with attempts to get score predictions from DOIs or full text URLs. In both cases, the results did not change the initial setup and evaluations were based solely on the format of the results and not their accuracy.

The main system prompt used was like that used in the previous paper and consists of the REF guidelines for the research area (Main Panel D) containing the information science in the REF (REF, 2019). The full prompt (Strategy 6) is in Appendix 1. The main changes are in the words at the start to cast it as an instruction to ChatGPT rather than information for REF assessors. The changes were motivated by prior research about prompting strategies (Yang et al., 2024) and an expectation that language similar to that reported by OpenAI in its documentation would be the most effective because it might reflect the format of the instructions fed to ChatGPT in its instruction following training phase.

Each score request was a single API call, specifying a ChatGPT model, including the system instructions, as in Appendix 1, and with the prompt, "Score the following journal article: ", followed by the article title/abstract/truncated text, as relevant. The maximum temperature parameter was set to 1, the default, the top_p parameter was also set to its default of 1, and the max_tokens parameter was set to 1000, which seemed adequate for the typical reports written by ChatGPT in the previous study (Thelwall, 2024).

## 2.3   GPT system prompt variations

Six variations of the basic REF prompt were tested to assess whether alternative prompts might give better results. In particular, the basic system prompt is lengthy and complex so it seems plausible that a simpler system prompt would be more effective. For this purpose, five truncated versions of the basic prompt were created by cutting it down at natural points (Appendix 1). A seventh prompt was created that did not request a score justification (Appendix 2) to test whether simpler results could be safely obtained from ChatGPT in contexts when only the score would be needed.

## 2.4   GPT score extraction

Perhaps because of the length of the REF guidelines in the system prompt, I did not find a way to get ChatGPT to report the score predictions systematically. Thus, each report contained an analysis of the article that (usually) contained a score assessment within it in different formats. Each ChatGPT model used a wide variety of structures to report its assessment and the models tested had their own structures and styles to some extent. A program was written to use all the patterns found to extract the score from the reports (Webometric Analyst: AI|ChatGPT: extract REF scores from report). For example, one rule was to extract the number between "Overall Score**: **" and "*". When three separate scores were given for rigour, originality, and significance, but not an overall score, these three scores were averaged to three decimal places. When ChatGPT reported a score range of two (usually, e.g., 3* to 4*) or three (occasionally, e.g., between 2* and 4*) the midpoint was taken. When the rules failed to find a score, the program prompted human input (me) to enter the score. In some cases (usually for titles or abstracts), scores were given for originality and significance but not for rigour. In such cases, the average of the originality and significance was used.

For some ChatGPT reports on titles and once for reports on abstracts, it did not give a score, instead reporting that it had insufficient information to make a judgement. These reports were registered as missing and ignored for the analysis. For example, if two of the 30 scores for article 1 were missing, then the article 1 average score would be the average of the remaining 28.

## 2.5   Analysis

Spearman correlations were used to assess the extent to which the 51 human scores agreed with the ChatGPT scores, without assuming that they are on the same scale. This assesses the extent to which the ChatGPT scores are in the same rank order as the human scores.

Since, as expected from a single previous experiment with the ChatGPT web interface (Thelwall, 2024), the scores for the same article differed between iterations, averaging them with the arithmetic mean may provide more stable predictions. To test the averaging, it seems reasonable to assume that the order in which the 30 iterations for each experiment was conducted would be irrelevant to the score. To assess the likely range of correlations for any number of ChatGPT iterations, confidence intervals and means were calculated for their correlations with human scores. For optimal results, different strategies were needed, depending on the number of iterations averaged (Webometric Analyst: Tab-sep|Stats|Randomly average 2-n columns and correlate with col n+1):

- 1 iteration: Correlations were calculated between each ChatGPT iteration and human scores, and the 30 correlations were averaged. The standard deviation was calculated to estimate confidence intervals for the mean correlation from a single iteration with the t distribution.
- Averaging 29 iterations: Correlations were calculated between the human scores and the average of 29 ChatGPT iterations (for each unique set of 29 out of 30 iterations) and the 30 correlations were averaged. The standard deviation was calculated to estimate confidence intervals with the t distribution.
- Averaging 2 or 28 iterations. In both cases there are 30x29=870 permutations of sets of iterations to average, so all 870 correlations were calculated, and the mean and standard deviation calculated for the average and confidence intervals.
- Averaging 3-27 iterations. In these cases, there are over 30x29x28=87028 permutations of sets of iterations to average, so a random set of 1000 permutations was selected (with replacement) and correlations were calculated for these. The mean and standard deviation was again calculated for the average and confidence intervals.
- Averaging all 30 iterations. In this case no permutations were possible and so no confidence intervals could be calculated. It is possible to calculate confidence intervals through bootstrapping with replacement, but this seems unnecessarily complex.

The accuracy of the predictions was assessed with Mean Absolute Deviation (MAD) to record the average difference between the predictions and human scores. Since model scores are not necessarily on the same scale as the human scores, linear regression models were fitted to transform them to the same scale in the simplest way. MAD calculations on the linear regression predictions therefore give a better estimate of the underlying accuracy of the predictions. The software used for this is correlation_and_regression.py in https://github.com/MikeThelwall/Python_misc.

# 3   Results

## 3.1   Input and averaging length comparisons: ChatGPT 3.5-turbo, 4o, and 4o-mini on truncated texts, abstracts, and titles

The ChatGPT reports were always well structured and usually plausible, often giving explicit reasons for the scores or making general statements like, "Rigour: [description]: this aligns with the criteria for a 3*". Reports on the titles alone were often short and without a prediction, stating instead that it was impossible to evaluate an article based on its title. For example, "Since the actual content and impact of the article [] is unknown, this evaluation remains speculative" (ChatGPT 4). These title-based reports were sometimes slightly inconsistent in that they would make conditional statements in the early parts, such as "If the methods align with expectations, then []" but in later parts give a more definitive evaluation, such as "Thus, this article is assigned a score of 3*".

In terms of the report scores, averaging more iterations of ChatGPT 3.5-turbo asymptotically increases the correlation with human scores, whatever the input (Figure 1). The rate of score increases tends to diminish as the number of iterations increases.

The optimal input for ChatGPT 3.5-turbo seems to be article abstracts (with titles) (Figure 1). Predictions based on these have the highest correlation with human scores. Extending abstracts to truncated full texts decreases the correlation but restricting the input text to article titles substantially decreases it. Nevertheless, score predictions from titles alone are surprisingly non-trivial, having a 0.46 correlation with human scores despite this minimal input.

LLMs all have limited context windows, but this does not seem to be the reason why ChatGPT 3.5 seems to perform better on shorter text. This may be because the full text was within the ChatGPT 3.5 context window in all cases. Nevertheless, it is possible that performance degrades for long text in some contexts even if it keeps within the context window (Hsieh et al., 2024). Nevertheless, both 4o and 4o-mini, with much longer context windows, also performed better on shorter text, although the difference was marginal for 4o (Table 1).

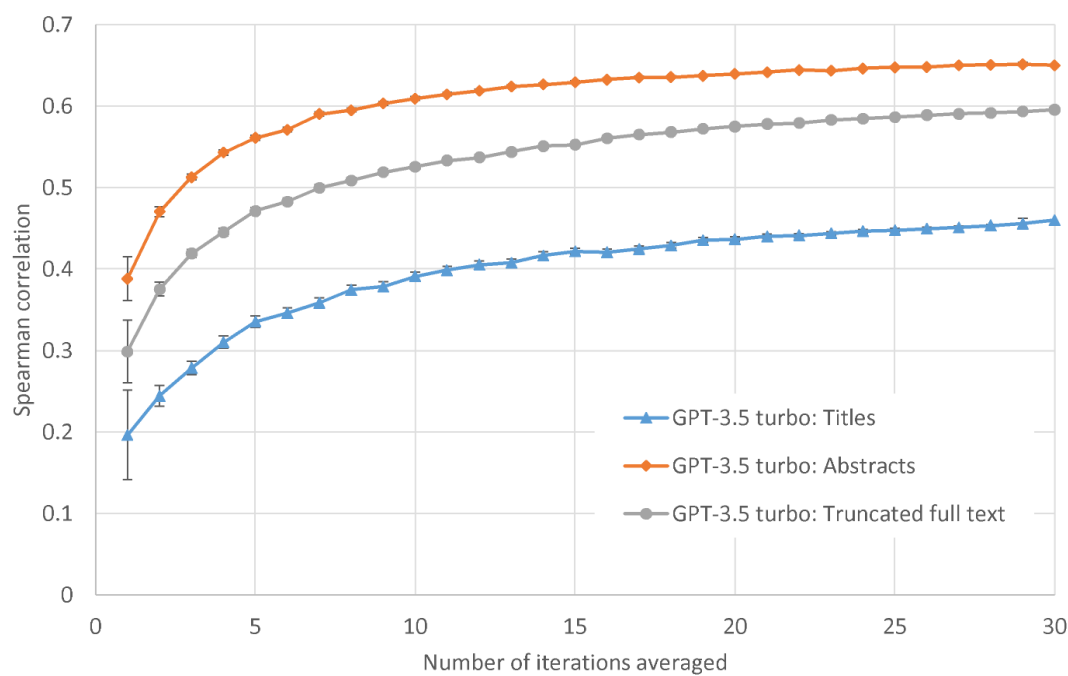

Figure 1. ChatGPT 3.5-turbo score prediction correlations against human scores for 51 information science article full texts (truncated), article titles and abstracts, or just titles. Averages over n iterations and confidence intervals are calculated as in the methods.

The predictions from the three types of input have varying correlations between themselves (rows and columns 3-5 of Table 1). The high correlation between the predictions based on titles and abstracts and the predictions based on truncated full texts (0.754) suggests that the presence of full text makes little difference.

Table 1. Spearman correlations between humans scores and model average scores (over 30 iterations) for 51 information science articles. Values above 0.75 are highlighted.

| Spearman correlation | GPT-3.5 turbo: Abstracts | GPT-3.5 turbo: Truncated text | GPT-4o-mini: Abstracts | GPT-4o-mini: Truncated text | GPT-4o: Abstracts | GPT-4o: Truncated text | Human |
|---|---|---|---|---|---|---|---|
| GPT-3.5 turbo: Titles | 0.439 | 0.444 | 0.359 | 0.499 | 0.539 | 0.589 | 0.434 |
| GPT-3.5 turbo: Abstracts | 1.000 | 0.757 | 0.700 | 0.718 | 0.875 | 0.774 | 0.674 |
| GPT-3.5 turbo: Truncated text | | 1.000 | 0.672 | 0.686 | 0.732 | 0.783 | 0.625 |
| GPT-4o-mini: Abstracts | | | 1.000 | 0.608 | 0.729 | 0.653 | 0.571 |
| GPT-4o-mini: Truncated text | | | | 1.000 | 0.813 | 0.801 | 0.506 |
| GPT-4o: Abstracts | | | | | 1.000 | 0.858 | **0.678** |
| GPT-4o: Truncated text | | | | | | 1.000 | 0.675 |

There is a tendency for score predictions to be higher with more input for the same model (Table 2), perhaps with the extra detail giving the model the extra evidence needed for a higher grade. This conjecture is supported by some of the reports citing a lack of evidence as a reason for not giving a higher score.

Table 2. Average humans scores and model average scores.

| | Human | GPT-3.5 turbo: Titles | GPT-3.5 turbo: Abstracts | GPT-3.5 turbo: Truncated text | GPT-4o-mini: Abstracts | GPT-4o-mini: Truncated text | GPT-4o: Abstracts | GPT-4o: Truncated text |
|---|---|---|---|---|---|---|---|---|
| Mean score | 2.75 | 2.49 | 2.75 | 3.03 | 2.93 | 3.22 | 2.99 | 3.16 |

## 3.2  Model comparison: ChatGPT 3.5-turbo, ChatGPT 4o and ChatGPT 4o-mini on abstracts

Unsurprisingly, the most powerful model (ChatGPT 4o) seems to give the best predictions but the difference between the models is not large. There is little difference between the relatively old ChatGPT 3.5-turbo and the new (at the time of writing) ChatGPT 4o-mini (Figure 2). Given that API calls with ChatGPT 4o are ten times more expensive than calls with ChatGPT 3.5-turbo and twenty times more expensive than ChatGPT 4o-mini calls (as of July 2024), this suggests that the cheaper models are good alternatives unless the highest accuracy is needed.

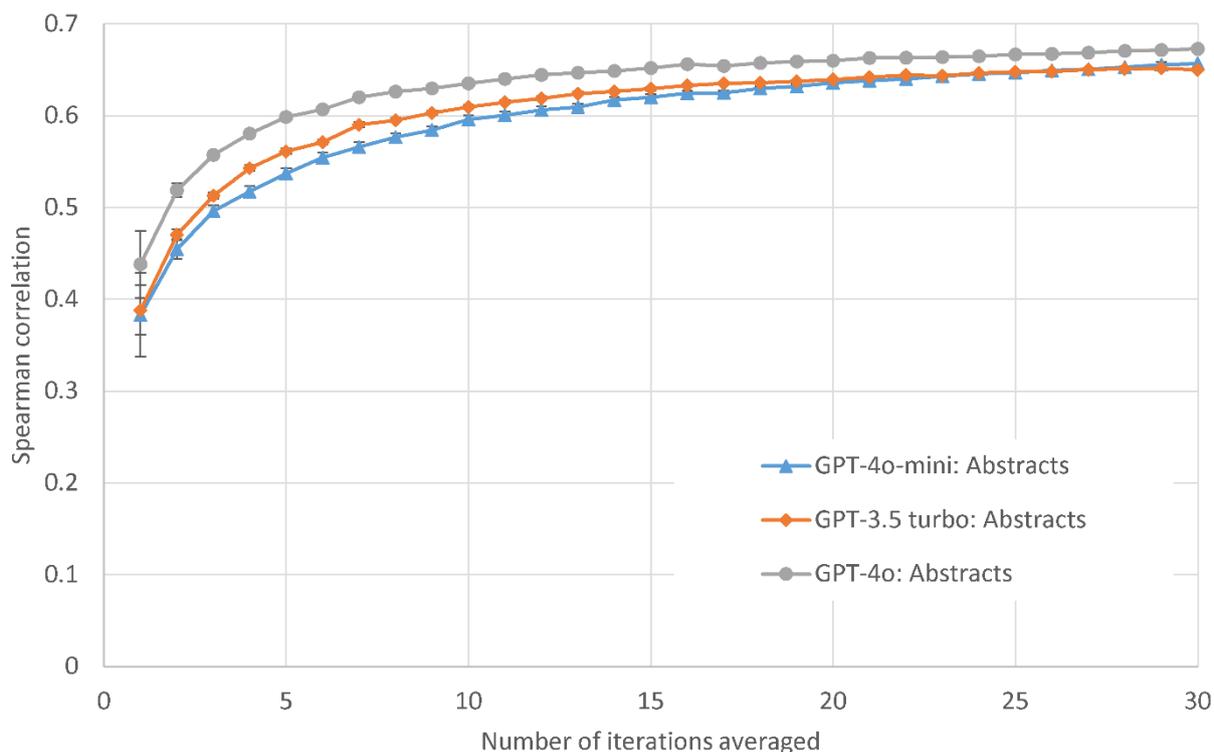

Figure 2. ChatGPT 4o-mini, ChatGPT 3.5-turbo and ChatGPT 4o score prediction correlations against human scores for 51 information science article titles and abstracts. Averages over n iterations and confidence intervals are calculated as in the methods.

A scatterplot of the ChatGPT predictions having the highest correlation with my scores is broadly consistent with a general trend for ChatGPT having some ability to distinguish between all four quality levels (Figure 3) rather than, for example, being able to distinguish 3* to 4* from lower quality research. There is too little data to provide strong evidence of this, however.

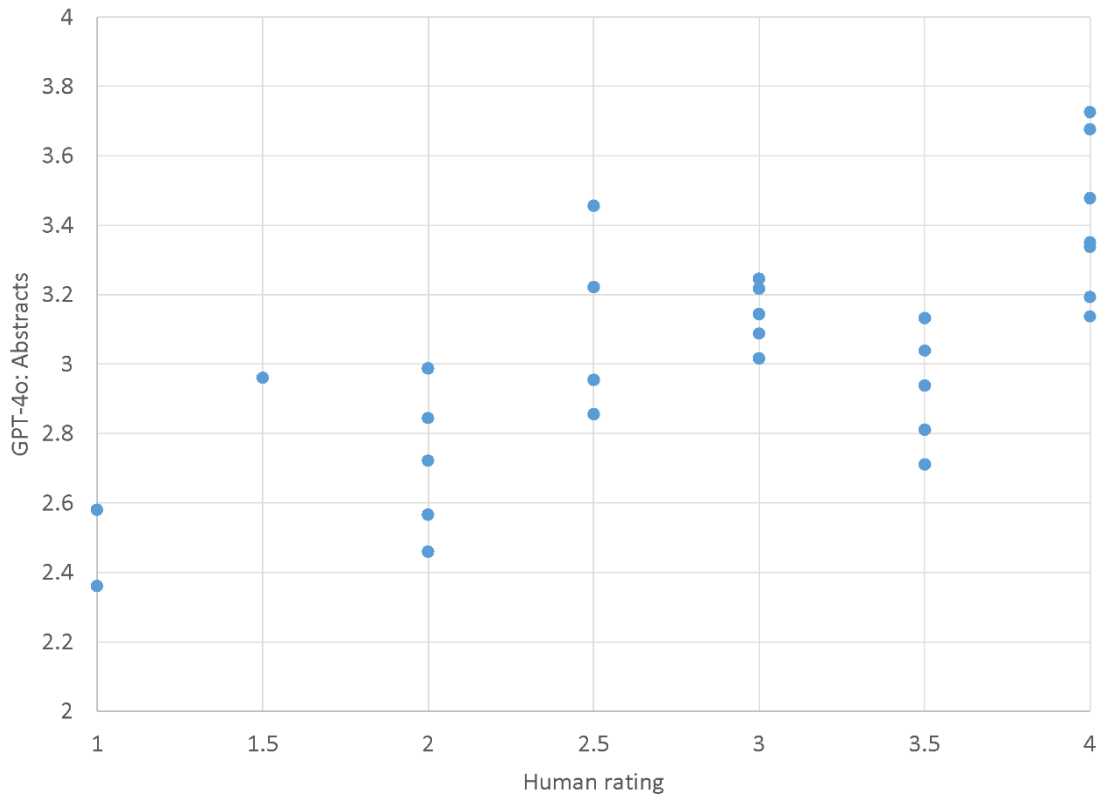

Figure 3. ChatGPT 4o score predictions based on abstracts (average of 30) against human scores (from the author) for 51 information science article titles and abstracts.

The predictions from the three models have even higher correlations between themselves than with the human scores (rows and columns 5-7 of Table 1). These very high correlations suggest that their language models have strong underlying similarities, despite the different formats of their reports.

### 3.3 Comparison of prompt strategies

Although all seven system prompts gave positive results, in the sense of correlations substantially above zero, the most complex strategy was the most effective on this dataset (Figure 4). By far the worst strategy was the one not requesting feedback in the ChatGPT report (Strategy 0), suggesting that asking for an analysis helps ChatGPT to judge scores better. This is despite ChatGPT occasionally giving its score before its analysis in response to other system prompts. The second-best system prompt is the single paragraph Strategy 2, and adding more information to it makes it worse (Strategies 3-5) until the quality level definitions are added (Strategy 6). If the Strategy 3 specific part (defining significance, rigour, originality) is removed from Strategy 6 then this performs slightly worse than strategy 6, however (correlation 0.641 after averaging 30) so the full instructions seem to be the optimal choice.

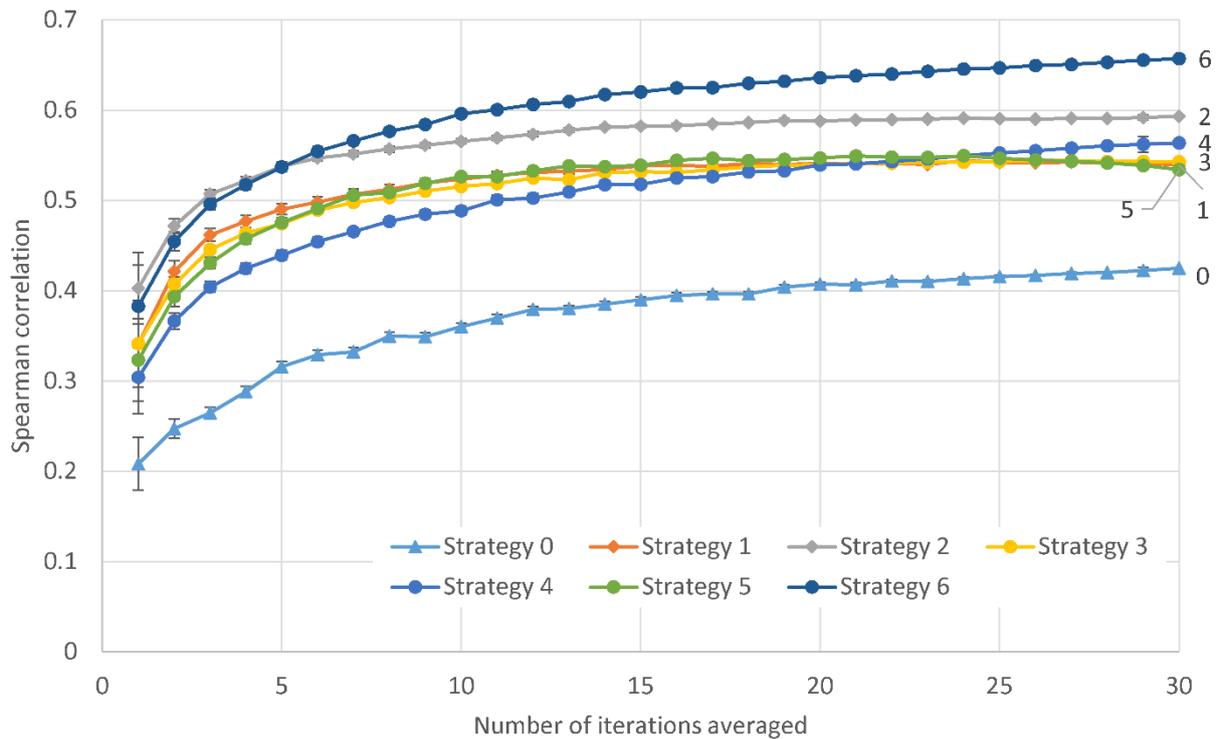

Figure 4. ChatGPT 4o score predictions based on abstracts (average of 30) against human scores (from the author) for 51 information science article titles and abstracts with seven different system prompts. Strategies 1-5 are abbreviations of Strategy 6, the full REF instructions, and Strategy 0 is a brief instruction without a request for justification.

### 3.4 Individual score level accuracy

Despite the positive correlations reported above, the prediction scores (Direct in Table 3) are, on average, closer to a wrong score than a correct score. They are nevertheless slightly better than the baseline strategy of assigning all articles the average score (2.75). The accuracy of the predictions can be strengthened with linear regression (Regression in Table 3) to the extent that they improve on the baseline strategy by up to 31% (i.e., they are 31% closer to the correct value than the 2.75 guesses are, on average). The regression MAD values are probably underestimates, however, due to overfitting – calculating accuracy on the data used to create the model.

The optimal prediction strategy in terms of accuracy for an article REF score in this topic area is therefore to run its title and abstract 30 times through GPT-4o, then multiply the average score by 2.05 and subtract 3.4. If the predicted scores need to be rounded to whole numbers, then the MADs do not change much (not shown in table), although the GPT-3.5 turbo: Abstracts MAD decreases from 0.51 to 0.45.

Table 3. Mean Average Deviations for direct predictions and predictions with linear regression for each model and input. The improve column gives the percentage reduction in MAD compared to the baseline strategy of assigning each article the overall human average, 2.75.

| | Direct | | Regression | | | |
|---|---|---|---|---|---|---|
| Model and input | MAD | Improve | Intercept | Coefficient | MAD | Improve |
| GPT-3.5 turbo: Titles | 0.68 | 6% | -1.16 | 1.57 | 0.63 | 13% |
| GPT-3.5 turbo: Abstracts | 0.60 | 17% | -3.46 | 2.26 | 0.51 | 30% |
| GPT-3.5 turbo: Truncated text | 0.70 | 4% | -7.49 | 3.38 | 0.55 | 24% |
| GPT-4o-mini: Abstracts | 0.63 | 13% | -3.32 | 2.07 | 0.59 | 19% |
| GPT-4o-mini: Truncated text | 0.75 | -3% | -2.44 | 1.61 | 0.60 | 17% |
| GPT-4o: Abstracts | 0.62 | 14% | -3.40 | 2.05 | 0.50 | 31% |
| GPT-4o: Truncated text | 0.69 | 5% | -4.44 | 2.28 | 0.50 | 31% |

## 4   Discussion

This study has many limitations, the most important of which is the restriction to a relatively small number of articles written by single person. The results could easily be substantially different for other fields and correlations are likely to be lower for sets of articles with a narrower quality range (probably including most REF submissions). Correlations might also be weaker for more mixed sets of articles, since LLMs might find quantitative research easier to classify or they might just use different scales for different types of research.

More generally, the reason why ChatGPT has some ability to report on the quality of academic research is unclear in terms of the main inputs that it is leveraging. For example, ChatGPT might have learned that some of the articles had attracted attention online or were associated with highly regarded journals even if only fed with an article's title and abstract. More research is needed to gain evidence of the factors used by LLMs to make research quality evaluations.

The results may also be better with future, larger models or with more precise system instructions than those listed in the Appendices. The results could also be better with other ChatGPT temperature and top_p parameters (which control different aspects of randomness), although the performance did not improve (for the abstracts dataset with GPT-4o-mini) if temperature was set to 0.1,0.5, 1.5 or 2, or if top_p was set to 0.25, 0.5, or 0.75. Finally, alternative current LLMs, such as Claude, Llama and the multimodal Google Gemini, might perform better.

### 4.1   Comparisons with prior work

The results improve on the prior attempt to predict REF scores on the same set of articles with ChatGPT 4 using the web interface and uploading PDFs or Word documents (Thelwall, 2024). Although this prior paper only averaged over 15 iterations, a graph of its results (Figure 5) shows that it is less accurate at this point than the three models tested in the current paper with abstracts as inputs (Figure 2) and it is also less accurate than the weaker ChatGPT 3.5-turbo with either abstracts or truncated full texts as inputs. This tends to confirm that inputting full text documents is a relatively ineffective approach for obtaining quality scores

The results also confirm with each dataset (e.g., Figures 1, 2, 3) the suggestion from a single previous experiment (Thelwall, 2024) that averaging over multiple iterations is a substantially more effective strategy than single runs of ChatGPT. This averaging produces different shape variations in graphs that all approximate the right-hand side of a sigmoid graph (the upper asymptote), with each additional iteration improving the results but adding less than the one before. In some cases (Figures 2, 4) the lines on the graphs cross, revealing small differences in shape. Thus, there is not an obvious maximum number of iterations that should normally be used, but five might be taken as sufficient for general purposes but additional iterations used when more precise estimates are needed (answering RQ2). Moreover, whilst ChatGPT can ingest documents that mix text and images, this comparison suggests that it is not powerful enough to leverage this capability for quality score prediction.

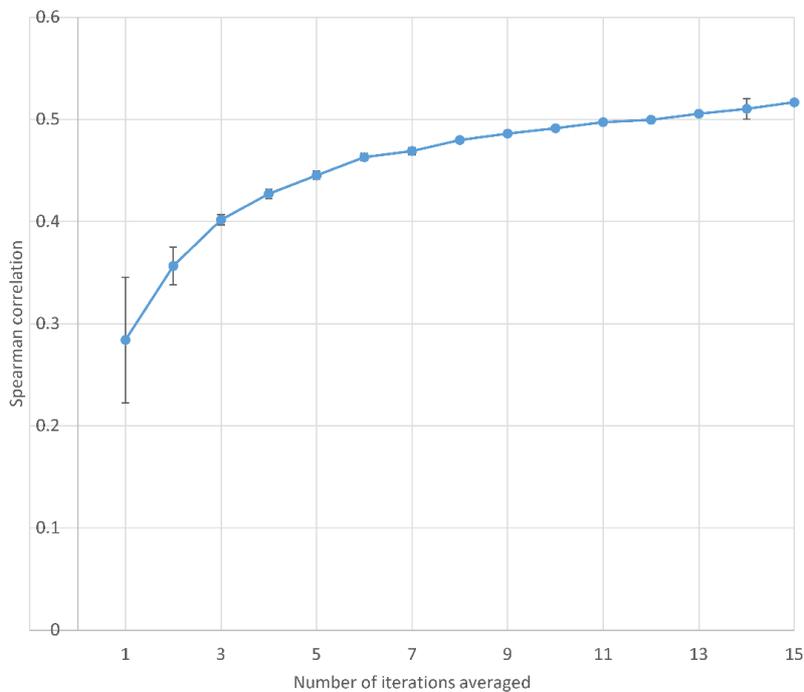

Figure 5. ChatGPT 4 (web interface) score prediction correlations against human scores for 51 information science article titles and abstracts. Averages over n iterations and confidence intervals are calculated as in the methods (data from: Thelwall, 2024).

The results thus agree with the prior study of different inputs that abstracts are better inputs than full texts (without images) (Zhou et al., 2024), and extend it by showing that this can also be true for post-publication journal article quality review (information science), in addition to pre-publication conference paper review (natural language processing). The current results also extend the prior study by showing that abstracts are better inputs than titles alone and (in combination with: Thelwall, 2024) better inputs than full texts (PDF or Word documents). This could be due to abstracts focusing on the core contributions of an article, whereas full texts give many small details that are less relevant, such as methods details. Thus, full texts as inputs may tend to obscure the main value of an article by including less relevant text that an LLM could pay partial attention to.

The correlation between the predictions and human estimates of research quality is substantially higher than found in the largest prior study of this relationship that used a machine learning approach. This prior study reported a Pearson correlation between machine learning REF quality predictions and REF scores of only 0.084 for Unit of Assessment 34, which contains information science. The highest correlation between the machine learning predictions and the scores was 0.562, for the Clinical Medicine Unit of Assessment (Thelwall et al., 2023), so ChatGPT has outperformed the state-of-the-art machine learning approach, despite only accessing a subset of its inputs. The prior traditional study incorporated journal, article and author team citation information as inputs in addition to frequent title and abstract terms, bigrams and trigrams. Although direct comparisons of the correlations are not fair because the REF quality scores are probably more concentrated, which tends to reduce correlations, this still suggests that ChatGPT is more promising than machine learning with bibliometric inputs for REF score prediction, albeit based on a single small case study.

The necessity for a complex prompt strategy does not seem to have a parallel finding in prior research. Nevertheless, the weak performance of the Strategy 0 system prompt partially aligns with prior research showing the effectiveness for complex tasks of "Let's think step by step" Chain-of-thought prompts that encourage an LLM to work through preliminary reasoning to get to an answer rather than guessing immediately (Zhang et al., 2022).

## 4.2 *Alternative input strategies*

In theory, it might also be possible to request a review of an article from ChatGPT by referencing the article's DOI or by submitting an URL of a public archive page or a free text online PDF. For this, ChatGPT API access does not allow content to be fetched from the web, but the ChatGPT web interface might. Experiments at the start of this project with these were unsuccessful. Web ChatGPT either reported that it could not find the document at the URL or wrote a plausible report on an imaginary article, sometimes citing the correct URL. This suggests currently that pointing ChatGPT to an article online is insufficient for an analysis and that it relies on the information being directly entered.

For example, using the web interface and ChatGPT 4o, the prompt "Score the following journal article: https://arxiv.org/abs/2104.09138" (an URL for the paper, "Cures, Treatments and Vaccines for Covid-19: International differences in interest on Twitter" assessed an unrelated article ("EfficientNet: Rethinking Model Scaling for Convolutional Neural Networks"), probably picking this article as a citation classic and therefore frequently met in its database. Repeating the same query produced assessments of similarly unrelated articles (e.g., "Electronically Programmable Photonic Molecule", "High-Resolution Image Synthesis with Latent Diffusion Models"). One of the repeated attempts identified the correct article, but did not write a report about it, and a follow-up identical query again identified the correct article and wrote a report about it. Thus, using URLs instead of article information is unreliable.

Following the above, ChatGPT 4o could be encouraged to hallucinate by entering partial titles. For example, the query "Score the following journal article: An experiment to ..." generated a fictional article title (ChatGPT 4o: "Assuming the journal article is titled "An Experiment to Improve Teaching Methods in Higher Education":") and a report evaluating it at 3* (ChatGPT 4: "This article could be rated 3* because it is internationally excellent in terms of originality, significance, and rigour but might fall short of the highest standards needed for a 4* rating due to limited global adoption or transformative impact."). Thus, entering short titles has additional risks.

## 5   Conclusion

Albeit on a single limited sample, the results suggest that the optimal strategy for estimating the REF-like research quality of an academic journal article is to use a system prompt derived from the unabbreviated REF2021 instructions, then feed the article title and abstract into ChatGPT 4o thirty times, and then apply a regression formula to transform it into the human scale. This gives a prediction that is on the same scale as the human evaluator and may have a high rank correlation with it (0.673 in the example here). The prediction is still inaccurate, however, being likely to deviate from the true score by 0.5, on average, despite the limited range of 1* to 4*. Others wishing to use this approach should calibrate the linear regression equation to their own quality standards by repeating the above experiment with 50+ articles that they have scored. Of course, article titles and abstracts can only be uploaded to ChatGPT if copyright permits. The web version of ChatGPT might use the input for training, so the copyright must allow this, and this probably excludes uploading CC BY copyright articles since ChatGPT will not generate copyright statements when the model trained on the data creates new content.

The ChatGPT accuracy, at the level of correlation, is the highest yet found for a quality prediction (e.g., Thelwall et al., 2023; Thelwall, 2024) and has the advantage of not requiring citation information. Nevertheless, at the individual article level, this is unacceptably inaccurate for important applications. In particular, it should not be used for peer review of conference papers or journal articles and also not for promotion and hiring decisions. It might be plausible to use average scores to compare sets of articles, as in the REF, or for comparing departments, if this did not introduce systematic biases (e.g., against qualitative research). More analysis is needed to understand how the predictions are made by ChatGPT to give insights into the types of biases that are likely.

# 6 References

Bang, Y., Cahyawijaya, S., Lee, N., Dai, W., Su, D., Wilie, B., & Fung, P. (2023). A multitask, multilingual, multimodal evaluation of ChatGPT on reasoning, hallucination, and interactivity. arXiv preprint arXiv:2302.04023.

Barrere, R. (2020). Indicators for the assessment of excellence in developing countries. In: Kraemer-Mbula, E., Tijssen, R., Wallace, M. L., & McClean, R. (Eds.), Transforming research excellence: New ideas from the Global South. Cape Town, South Africa: African Minds (pp. 219-232).

Bornmann, L., & Lepori, B. (2024). The use of ChatGPT to find similar institutions for institutional benchmarking. Scientometrics, 1-6.

Buckle, R. A., and Creedy, J. (2024). The performance based research fund in New Zealand: Taking stock and looking forward. New Zealand Economic Papers, 58(2), 107-125. https://doi.org/10.1080/00779954.2022.2156382

Buscemi, A., & Proverbio, D. (2024). Chatgpt vs Gemini vs Llama on multilingual sentiment analysis. arXiv preprint arXiv:2402.01715.

Carbonell Cortés, C. (2024). AI-assisted pre-screening of biomedical research proposals: ethical considerations and the pilot case of "la Caixa" Foundation. https://www.youtube.com/watch?v=O2DcXzEtCmg

de Winter, J. (2024). Can ChatGPT be used to predict citation counts, readership, and social media interaction? An exploration among 2222 scientific abstracts. Scientometrics, 1-19.

Elsevier (2024). Publishing ethics. https://www.elsevier.com/en-gb/about/policies-and-standards/publishing-ethics (20 July 2024)

Hsieh, C. P., Sun, S., Kriman, S., Acharya, S., Rekesh, D., Jia, F., & Ginsburg, B. (2024). RULER: What's the Real Context Size of Your Long-Context Language Models? arXiv preprint arXiv:2404.06654.

Kang, D., Ammar, W., Dalvi, B., Van Zuylen, M., Kohlmeier, S., Hovy, E., & Schwartz, R. (2018). A dataset of peer reviews (PeerRead): Collection, insights and NLP applications. arXiv preprint arXiv:1804.09635.

Kocoń, J., Cichecki, I., Kaszyca, O., Kochanek, M., Szydło, D., Baran, J., & Kazienko, P. (2023). ChatGPT: Jack of all trades, master of none. Information Fusion, 101861.

Kousha, K., & Thelwall, M. (2024). Factors associating with or predicting more cited or higher quality journal articles: An Annual Review of Information Science and Technology (ARIST) paper. Journal of the Association for Information Science and Technology, 75(3), 215-244.

Langfeldt, L., Nedeva, M., Sörlin, S., & Thomas, D. A. (2020). Co-existing notions of research quality: A framework to study context-specific understandings of good research. Minerva, 58(1), 115-137.

Liang, W., Izzo, Z., Zhang, Y., Lepp, H., Cao, H., Zhao, X., & Zou, J. Y. (2024a). Monitoring ai-modified content at scale: A case study on the impact of ChatGPT on AI conference peer reviews. arXiv preprint arXiv:2403.07183.

Liang, W., Zhang, Y., Cao, H., Wang, B., Ding, D. Y., Yang, X., & Zou, J. (2024b). Can large language models provide useful feedback on research papers? A large-scale empirical analysis. *NEJM AI*, AIoa2400196. https://doi.org/10.1056/AIoa2400196

Lu, Y., Xu, S., Zhang, Y., Kong, Y., & Schoenebeck, G. (2024). Eliciting Informative Text Evaluations with Large Language Models. arXiv preprint arXiv:2405.15077.

MacRoberts, M. H., & MacRoberts, B. R. (2018). The mismeasure of science: Citation analysis. Journal of the Association for Information Science and Technology, 69(3), 474-482.

Moed, H. F. (2006). Citation analysis in research evaluatio). Berlin, Germany: Springer.

OpenAI (2024). Key concepts. https://platform.openai.com/docs/concepts (21 July 2023).

Pontika, N., Klebel, T., Correia, A., Metzler, H., Knoth, P., & Ross-Hellauer, T. (2022). Indicators of research quality, quantity, openness, and responsibility in institutional review, promotion, and tenure policies across seven countries. Quantitative Science Studies, 3(4), 888-911.

Qiu, J., & Han, X. (2024). An Early Evaluation of the Long-Term Influence of Academic Papers Based on Machine Learning Algorithms. IEEE Access, 12, 41773-41786.

REF (2019). Panel criteria and working methods (2019/02). https://2021.ref.ac.uk/publications-and-reports/panel-criteria-and-working-methods-201902/index.html

Sivertsen, G. (2017). Unique, but still best practice? The Research Excellence Framework (REF) from an international perspective. Palgrave Communications, 3(1), 1-6.

Thelwall, M. (2024). Can ChatGPT evaluate research quality? *Journal of Data and Information Science*, 9(2), 1-21. https://doi.org/10.2478/jdis-2024-0013

Thelwall, M., Kousha, K., Wilson, P. Makita, M., Abdoli, M., Stuart, E., Levitt, J., Knoth, P., & Cancellieri, M. (2023). Predicting article quality scores with machine learning: The UK Research Excellence Framework. Quantitative Science Studies, 4(2), 547-573. https://doi.org/10.1162/qss_a_00258

Tierney, W. G., & Bensimon, E. M. (1996). Promotion and tenure: Community and socialization in academe. New York, NY: Suny Press.

Yang, J., Jin, H., Tang, R., Han, X., Feng, Q., Jiang, H., & Hu, X. (2024). Harnessing the power of llms in practice: A survey on ChatGPT and beyond. ACM Transactions on Knowledge Discovery from Data, 18(6), 1-32.

Zhang, Z., Zhang, A., Li, M., & Smola, A. (2022). Automatic chain of thought prompting in large language models. arXiv preprint arXiv:2210.03493.

Zhou, R., Chen, L., & Yu, K. (2024). Is LLM a Reliable Reviewer? A Comprehensive Evaluation of LLM on Automatic Paper Reviewing Tasks. *LREC-COLING 2024* (pp. 9340-9351).

# 7 Appendix 1: System prompts for ChatGPT (adapted from: REF, 2019)

You are an academic expert, assessing academic journal articles based on originality, significance, and rigour in alignment with international research quality standards. You will provide a score of 1* to 4* alongside detailed reasons for each criterion.[Strategy 1 is everything before here] You will evaluate innovative contributions, scholarly influence, and intellectual coherence, ensuring robust analysis and feedback. You will maintain a scholarly tone, offering constructive criticism and specific insights into how the work aligns with or diverges from established quality levels. You will emphasize scientific rigour, contribution to knowledge, and applicability in various sectors, providing comprehensive evaluations and detailed explanations for its scoring.

[Strategy 2 is everything above here]

Originality will be understood as the extent to which the output makes an important and innovative contribution to understanding and knowledge in the field. Research outputs that demonstrate originality may do one or more of the following: produce and interpret new empirical findings or new material; engage with new and/or complex problems; develop innovative research methods, methodologies and analytical techniques; show imaginative and creative scope; provide new arguments and/or new forms of expression, formal innovations, interpretations and/or insights; collect and engage with novel types of data; and/or advance theory or the analysis of doctrine, policy or practice, and new forms of expression.

Significance will be understood as the extent to which the work has influenced, or has the capacity to influence, knowledge and scholarly thought, or the development and understanding of policy and/or practice.

Rigour will be understood as the extent to which the work demonstrates intellectual coherence and integrity, and adopts robust and appropriate concepts, analyses, sources, theories and/or methodologies.

[Strategy 3 is everything above here]

The scoring system used is 1*, 2*, 3* or 4*, which are defined as follows.

4*: Quality that is world-leading in terms of originality, significance and rigour.

3*: Quality that is internationally excellent in terms of originality, significance and rigour but which falls short of the highest standards of excellence.

2*: Quality that is recognised internationally in terms of originality, significance and rigour.

1* Quality that is recognised nationally in terms of originality, significance and rigour.

[Strategy 4 is everything above here]

The terms 'world-leading', 'international' and 'national' will be taken as quality benchmarks within the generic definitions of the quality levels. They will relate to the actual, likely or deserved influence of the work, whether in the UK, a particular country or region outside the UK, or on international audiences more broadly. There will be no assumption of any necessary international exposure in terms of publication or reception, or any necessary research content in terms of topic or approach. Nor will there be an assumption that work published in a language other than English or Welsh is necessarily of a quality that is or is not internationally benchmarked.

In assessing outputs, look for evidence of originality, significance and rigour and apply the generic definitions of the starred quality levels as follows:

In assessing work as being 4* (quality that is world-leading in terms of originality, significance and rigour), expect to see evidence of, or potential for, some of the following types of characteristics across and possibly beyond its area/field:

a primary or essential point of reference;

of profound influence;

instrumental in developing new thinking, practices, paradigms, policies or audiences;

a major expansion of the range and the depth of research and its application;

outstandingly novel, innovative and/or creative.

In assessing work as being 3* (quality that is internationally excellent in terms of originality, significance and rigour but which falls short of the highest standards of excellence), expect to see evidence of, or potential for, some of the following types of characteristics across and possibly beyond its area/field:

an important point of reference;

of considerable influence;

a catalyst for, or important contribution to, new thinking, practices, paradigms, policies or audiences;

a significant expansion of the range and the depth of research and its application;

significantly novel or innovative or creative.

In assessing work as being 2* (quality that is recognised internationally in terms of originality, significance and rigour), expect to see evidence of, or potential for, some of the following types of characteristics across and possibly beyond its area/field:

a recognised point of reference;

of some influence;

an incremental and cumulative advance on thinking, practices, paradigms, policies or audiences;

a useful contribution to the range or depth of research and its application.

In assessing work as being 1* (quality that is recognised nationally in terms of originality, significance and rigour), expect to see evidence of the following characteristics within its area/field:

an identifiable contribution to understanding without advancing existing paradigms of enquiry or practice;

of minor influence.

# 8   Appendix 2: System prompt Strategy 0 (adapted from: REF, 2019)

You are an academic research expert. Your job is to assign research quality scores to journal articles on a scale of 1* to 4*.

The scoring system used is 1*, 2*, 3* or 4*, which are defined as follows.

4*: Quality that is world-leading in terms of originality, significance and rigour.

3*: Quality that is internationally excellent in terms of originality, significance and rigour but which falls short of the highest standards of excellence.

2*: Quality that is recognised internationally in terms of originality, significance and rigour.

1* Quality that is recognised nationally in terms of originality, significance and rigour.